\documentclass[prc,aps,superscriptaddress,showpacs,nofootinbib,twocolumn]{revtex4}
\usepackage{amssymb}
\usepackage[dvips]{graphicx}
\usepackage[english]{babel}
\usepackage{indentfirst}
\usepackage{amsxtra}
\usepackage{amsmath}
\usepackage{supertabular}
\usepackage{multirow}
\usepackage [mathcal]{eucal}
\begin{document}
\title {\bf Pair-transfer probability in open- and closed-shell Sn isotopes}

\author{M. Grasso}
\affiliation{Institut de Physique Nucl\'eaire, IN2P3-CNRS, Universit\'e Paris-Sud, 
F-91406 Orsay Cedex, France}

\author{D. Lacroix}
\affiliation{Grand Acc\'el\'erateur National d'Ions Lourds (GANIL), CEA/DSM-CNRS/IN2P3, Bvd Henri Becquerel, F-14076 Caen, France}

\author{A. Vitturi}
\affiliation{Dipartimento di Fisica G. Galilei, via Marzolo 8, I-35131 Padova, Italy}
\affiliation{Istituto Nazionale di Fisica Nucleare (INFN), Sezione di Padova, via Marzolo 8, I-35131 Padova, Italy}

\begin{abstract} 

Approximations made to estimate two-nucleon transfer probabilities in ground-state to ground-state transitions and physical interpretation 
of these probabilities are discussed. 
Probabilities are often calculated by approximating both ground states, of the initial nucleus 
$A$ and of the final nucleus $A\pm2$ by the same quasiparticle vacuum. We analyze two 
improvements of this approach. First, the effect of using two different ground states with average numbers of particles $A$ and $A\pm2$ is quantified. Second, by using projection
techniques, the role of particle number restoration is analyzed.  Our analysis shows that 
the improved treatment plays a role close to magicity, leading to an enhancement of the pair-transfer 
probability. In mid-shell regions, part of the error made by approximating the initial and final ground states by a single vacuum is 
compensated by projecting onto good particle number. 
Surface effects are analyzed by using pairing interactions with a different 
volume-to-surface mixing. 
Finally, a simple expression 
of the pair-transfer probability is given in terms of occupation probabilities in the canonical basis. 
We show that, in the canonical basis formulation, surface effects which are visible in the transfer probability are related to the fragmentation of single-particle occupancies close to the Fermi energy. This provides a complementary interpretation with respect to the standard quasiparticle representation where surface effects are generated by the integrated radial profiles of the contributing wave functions. 
\end{abstract} 

\vskip 0.5cm  \maketitle 
%

\section{Introduction}

The link between Cooper-pair superfluidity in nuclei and cross sections associated to pair-transfer reactions has been extensively discussed in the literature starting from the early work of Broglia and collaborators in the 70s \cite{broglia}. In particular, the relation between the characteristics of the  pairing correlations  and the transfer probabilities in two-particle transfer reactions has been analyzed \cite{oertzen}. 
Recently, there is a renewal of interest on experimental 1-nucleon ($1n$), 2-nucleon ($2n$) and more generally multi-nucleon transfer channels at bombarding energies close to the Coulomb barrier \cite{Cor11}.
From the theoretical side, in the last decade, new microscopic calculations have been developed, essentially in the framework of the Hartree-Fock-Bogoliubov (HFB) + quasiparticle random-phase approximation (QRPA) theory to investigate the properties of the $0^+$ 
\cite{khan04} and $2^+$ \cite{Mat11} excitation modes associated to $2n$ addition or removal during transfer reactions. 
A similar analysis has been performed also within the time-dependent HFB 
model in the small-amplitude limit \cite{avez}. The possibility to use pairing vibrations to constrain the pairing interaction employed in HFB-based calculations has been explored recently \cite{khan,pllumbi}. The objective of these last studies was to suggest an experimental measurement adapted to identify the surface/volume mixing character of the pairing interaction. Two-neutron ($p,t$) transfer reactions in very neutron-rich Sn isotopes have been indicated as a good experimental candidate. 

In the recent Ref. \cite{matsuo}, the effects of the surface/volume nature of the pairing interaction on $2n$ transfer have been extensively analyzed. The enhancement of transfer probabilities at the surface has been predicted (when a surface-peaked interaction is used) in the transition from the ground state (GS)
of the nucleus with mass $A$ to the GS of the nuclei with masses 
 $A\pm2$ in Sn isotopes beyond $N=82$. Similarly to what done in other recent estimates of the pair-transfer probability using the microscopic HFB approach \cite{Cor11}, the strength associated to these transitions has been calculated in Ref. \cite{matsuo} with an approximate formula where only the wave functions of the nucleus $A$ enter. This approximate treatment differs from the original formulation given in Ref. \cite{broglia} where 
the components of both nuclei $A$ and $A\pm2$ appear. 
Accordingly, in other recent works not based on HFB \cite{potel}, in the expressions of the two-particle transfer spectroscopic amplitudes the wave functions of the two nuclei $A$ and $A+2$ appear  \cite{potel}. 
It is worth mentioning that, even the formulas given in Ref. \cite{broglia} are approximate expressions and, as far as we know, the underlying 
approximation needs to be clarified. 

One goal of the present work is to discuss different level of approximation used to 
estimate two-nucleon transition probabilities. 
In this work we derive the expressions to be used for the two-particle GS $ \rightarrow$ GS transfer probabilities in the framework of the HFB model. We compare our results with those obtained with the model of Shimoyama and Matsuo  \cite{matsuo} and analyze the differences in mid-shells and at shell closures. Following their work, different values for the surface/volume mixing parameter are used in the pairing interaction and application is made for the chain of  Sn isotopes. By using the canonical basis representation of the HFB model we provide a complementary interpretation of surface effects. Finally, we quantify the effect of particle number restoration on two-neutron transfer probabilities. 

The article is organized as follows. In Sec. II the general scheme of the present calculations is  presented. In Sec. III, an expression for 
the transfer probability is derived in an approximated framework analogous to that of Ref. \cite{matsuo} in the quasiparticle (coordinate) (Subsec. III-A) and in the canonical basis (Subsec. III-B) representation. The interpretation of the pair-transfer probability is discussed in Subsec. III-C. In Subsec. III-D the corresponding results are shown and compared. Improved formulas for the probability are derived in both the quasiparticle (Subsec. IV-A) and the 
canonical basis (Subsec. IV-B) formulations. Results are presented and commented in Subsec. IV-C. 
Finally, projection techniques are applied in the canonical basis case. 
The effect of particle number conservation on 
two-neutron transition probabilities is discussed (Sec. V). In Sec. VI, a summary is drawn and perspectives are outlined.  

\section{Framework of the present calculations}

The removal and addition GS $\rightarrow$ GS pair-transfer amplitudes are written as,  
\begin{equation}
T^{\rm Rem}_{\rm  GS} = \langle {\rm GS}_{A-2} | \Psi_{q'} ({\bf{r_1}},-\sigma_1) \Psi_q ({\bf{r_2}},\sigma_2) | {\rm GS}_{A} \rangle, 
\label{rem}
\end{equation}
and
\begin{equation}
T^{\rm  Add}_{\rm  GS} = \langle {\rm GS}_{A+2} | \Psi^{\dag}_{q'} ({\bf{r_2}},\sigma_2) \Psi^{\dag}_q ({\bf{r_1}},-\sigma_1) | {\rm GS}_{A} \rangle, 
\label{add}
\end{equation}
respectively, 
where $\sigma$ and $q$ represent spin and isospin, respectively. These matrices are assumed diagonal in isospin, that is $q'=q$. 
The states $| {\rm GS}_{A} \rangle$ and $| {\rm GS}_{A \pm 2} \rangle$ correspond a priori to the highly correlated ground states 
in the entrance and exit channel of the transfer reaction. For medium and heavy nuclei, only approximate wave functions can be 
obtained

Here, the framework of the Skyrme-HFB model with a zero-range density-dependent pairing interaction $V_{pair}$ is used with 
\begin{equation}\label{eq:vpair}
 V_{pair}({\bf r_1},{\bf r_2})=V_0\left[1-\eta\left(\frac{\rho({\bf R})}{\rho_0}\right)^\alpha\right]
 \delta\left({\bf r_1}-{\bf r_2}\right),
\end{equation}
where ${\bf R}=({\bf r_1}+{\bf r_2})/2$. The parameters $\alpha$ and $\rho_0$ are chosen equal to 1 and 0.16 fm$^{-3}$, respectively. With a cutoff equal to 60 MeV in quasiparticle energies and a maximum value of $j$ equal to 15/2, the parameter $V_0$ is adjusted to reproduce the two-neutron separation energies of Sn isotopes as already done in Refs. \cite{khan,pllumbi}. Two different pairing interactions are employed in terms of surface/volume mixing to check the sensitivity of the results on two different pairing radial localizations. The values of the parameter $V_0$ which are used here are  $V_0$ = -670 MeV fm$^{-1}$ for a pure surface interaction ($\eta=1$) and  $V_0$ = -390 MeV fm$^{-1}$ for a mixed interaction ($\eta=0.65$). The Skyrme interaction which is employed in the present calculations is SLy4 \cite{chabanat}.
The HFB calculations are performed in coordinate representation with a box discretization and a box radius equal to 20 fm. The pair-transfer probabilities are derived both in the quasiparticle framework and in the canonical basis. In the latter case, it will be shown that the integrals of the radial parts of the wave functions disappear in the expressions of the pair-transfer probabilities (which depend only on the occupation numbers).  This allows us to check whether the differences found with different pairing potential profiles (enhancement effects when a surface-peaked interaction is used) are artificial effects which are provided only by the formulations where integrals of wave functions are present or genuine physical effects that are found in all types of formulations.   

Besides the approximation made on the many-body wave function within the HFB approach, further approximations have to be done to estimate the pair-transfer probability. These approximations are analyzed in the following sections.

\section{Pair-transfer probabilities within an approximated scheme in the HFB theory}

\subsection{Quasiparticle formulation}

In Ref. \cite{matsuo} the removal and addition amplitudes are calculated by assuming that the ground state of the nucleus $A \pm 2$ can be approximated by the ground state of the nucleus $A$ in the evaluation of Eqs. (\ref{rem}) and (\ref{add}). By taking $\bf{r_1} = \bf{r_2} = \bf{r}$ and by assuming spherical symmetry, this approximation leads to  
\begin{eqnarray}
\nonumber
T^{\rm  Rem}_{\rm  GS}\ &\sim& T^{\rm  Add}_{\rm  GS} \sim \langle {\rm GS}_{A} | \Psi_{q} ({\bf{r}},-\sigma_1) \Psi_q ({\bf{r}},\sigma_2) | {\rm GS}_{A} \rangle \\
&=& - \frac{1}{4 \pi r^2} \sum_{nlj} (2j+1) u^A_{nlj} (r) v^A_{nlj} (r)= 
\kappa(r),  
\label{apprgs}
\end{eqnarray}
where $u^A(r)$ and $v^A(r)$ are the radial parts of the upper and lower components of the quasiparticle wave functions for the nucleus $A$, respectively, and $\kappa$ is the anomalous density of the nucleus $A$; the index $n$ runs over the number of states in each ($j,l$) channel. To derive the above expression in terms of the $u$ and 
$v$ components, the Bogoliubov transformations have been used (after having done the approximation $|GS_{A\pm2} \rangle \sim  |GS_{A} \rangle $), which can be written in the following way, 
\begin{equation}
 \Psi ({\bf r},\sigma) =\sum_n \left[ u^A_n ({\bf r} , \sigma) \gamma_{n \sigma} + (-1)^{1/2 + \sigma} v^{A*}_n
({\bf r} , -\sigma) \gamma^{\dagger}_{n -\sigma} \right].
\end{equation}

One may expect that the approximation leading to Eq. (4) is reasonable for mid-shell nuclei. However, at shell closures, the ground states of the nuclei $A$ and $A \pm 2$ are expected to be quite different 
and the anomalous density is zero  due to the collapse of static pairing correlations. 
It can thus be interesting to check especially in these regions the validity of such an  approximation.  

The pair-transfer probability calculated by using the amplitude given by Eq. (\ref{apprgs}) is written as
\begin{eqnarray}
P^{\rm  Rem}_{GS} (A)&=& P^{\rm  Add}_{GS} (A) \nonumber \\
&=& \left| \int dr \sum_{nlj} (2j+1) u^A_{nlj} (r) v^{A*}_{nlj} (r) \right|^2.  \label{eq:matsuo}
\end{eqnarray} 

\subsection{Canonical basis formulation}
\label{sec:can}

An alternative formulation can be developed in the canonical basis formalism. 
Let us introduce 
the canonical states, denoted by $\{\varphi_i \}$ associated to the set of creation/annihilation 
operators $\{a^\dagger_i , a_i \}$ where $i$ contains all quantum numbers.
The corresponding occupation number is denoted by $n^A_i$, where $A$ is the mass 
of the nucleus.
Accounting for the convention for the time-reversed states used in Ref. \cite{Dob96} we have 
\begin{eqnarray}
a^\dagger_i & = & \int d{\mathbf r}  \left\{ \varphi_i ({\bf r}, \uparrow)  \Psi^\dagger ({\mathbf r} ,\uparrow) +  
\varphi_i ({\bf r}, \downarrow) \Psi^\dagger({\mathbf r} ,\downarrow) \right\}, \nonumber \\
a^\dagger_{\bar i} & = &  
\int d{\mathbf r}  \left\{ \varphi^*_{i} ({\bf r}, \downarrow) \Psi^\dagger ({\mathbf r} ,\uparrow) -
\varphi^*_{i} ({\bf r}, \uparrow)  \Psi^\dagger({\mathbf r} ,\downarrow) \right\},
\end{eqnarray}      
where $\Psi$ are the operators introduced in the Bogoliubov transformations. 
These equation can eventually be inverted to give 
\begin{eqnarray}
\Psi^\dagger ({\mathbf r} ,\uparrow)  & = & \sum_{i>0} \left\{ a^\dagger_i  \varphi^*_i ({\bf r}, \uparrow) + a^\dagger_{\bar i} \varphi^*_{\bar i} ({\bf r}, \uparrow) \right\},   \\
\Psi^\dagger ({\mathbf r} ,\downarrow)  & = &  
\sum_{i>0} \left\{ a^\dagger_i  \varphi^*_i ({\bf r}, \downarrow) + a^\dagger_{\bar i} \varphi^*_{\bar i} ({\bf r}, \downarrow) \right\}.
\end{eqnarray} 
In the canonical basis, the HFB quasiparticle ground state takes a BCS-like form,
\begin{eqnarray}
| {\rm GS}_A \rangle   =  \prod_{i>0} \left( u^A_i + v^A_i a^\dagger_i a^\dagger_{\bar i} \right) | 0 \rangle, \label{eq:bcs} 
\end{eqnarray} 
where $u^A_i = \sqrt{1-n^A_i}$ and $v^A_i = \sqrt{n^A_i}$. 
In what follows, we consider the addition pair-transfer probability; the expression for the removal probability can be derived in     
the same way. We have:
 \begin{eqnarray}
T^{\rm Add}_{\rm GS} ({\bf r}) & = & \langle {\rm GS}_{A+2} | 
\sum_{i, j>0} \left\{ a^\dagger_j \varphi^*_j ({\bf r}, \uparrow) + a^\dagger_{\bar j} \varphi^*_{\bar j} ({\bf r}, \uparrow) \right\}
\nonumber \\
&&
\left\{ a^\dagger_i  \varphi^*_i ({\bf r}, \downarrow) + a^\dagger_{\bar i} \varphi^*_{\bar i} ({\bf r}, \downarrow) \right\}
 | {\rm GS}_A \rangle. 
\end{eqnarray}
Similarly to what is done in Ref. \cite{matsuo} and in the previous section, one can eventually assume that $\langle {\rm GS}_{A+2} | $ can be replaced 
by $\langle {\rm GS}_{A} | $. By taking advantage of the fact that the single-particle states are canonical and by using the 
properties relating time-reversed states, one finally deduces that   

\begin{widetext}

\begin{eqnarray}
\nonumber
T^{\rm Add}_{\rm GS} ({\bf r}) &=& T^{\rm Rem}_{\rm GS} ({\bf r}) 
 \simeq  - \sum_{i>0} \langle {\rm GS}_A | a^\dagger_i  a^\dagger_{\bar i} | {\rm GS}_A \rangle \left(|\varphi_i ({\bf r}, \uparrow) |^2  + |\varphi_i ({\bf r}, \downarrow) |^2 \right) \nonumber \\
&=& - \sum_{i>0} \sqrt{n^{A}_i (1 - n^{A}_i)}  \left(|\varphi_i ({\bf r}, \uparrow) |^2  + |\varphi_i ({\bf r}, \downarrow) |^2 \right). \label
 {eq:bcstrans}
\end{eqnarray}  
In the special case considered here, where the nucleus is assumed spherical, we finally obtain 
\begin{equation}
T^{\rm Add}_{\rm GS} ({\bf r}) 
 =  - \frac{1}{4\pi r^2} \sum_{nlj} (2j+1) \sqrt{n^{A}_{nlj} (1 - n^{A}_{nlj})} |\phi_{nlj} (r) |^2, 
\end{equation}
\end{widetext}  
where $\phi_{nlj}$ stands for the radial part of the canonical basis component.
This expression provides an alternative form of the transition amplitude and its numerical estimate should exactly match 
with the one obtained with Eq. (4).
Interestingly enough, in this specific basis, the radial dependence of the wave function completely disappears in the transition probability 
\begin{eqnarray}
P^{\rm Add}_{\rm GS} (A) &=&  P^{\rm Rem}_{\rm GS} (A) = \left|  \sum_i \sqrt{n^{A}_i (1 - n^{A}_i)} \right|^2, 
\end{eqnarray}    
or, equivalently, in spherical symmetry, 
\begin{eqnarray}
\nonumber
P^{\rm Add}_{\rm GS} (A)  &=& P^{\rm Rem}_{\rm GS} (A) \\ 
&=& \left|  \sum_{nlj} (2j+1) \sqrt{n^{A}_{nlj} (1 - n^{A}_{nlj})} \right|^2. \label{eq:pacan}
\end{eqnarray}

\subsection{Comparison between quasiparticle and canonical formulation}

\begin{figure}[htb]
\begin{center}
\includegraphics[width=8cm]{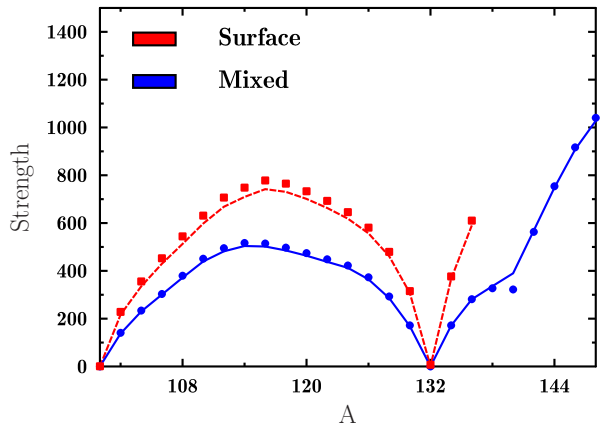}
\end{center}
\caption{Comparison of the removal (addition) probability obtained by using Eq. (\ref{eq:matsuo}) and  for  
the mixed pairing case (blue solid line) and the pure surface case (red dashed line). The results obtained using Eq. (\ref{eq:pacan})
are also shown for the mixed (filled circles) and pure surface (blue filled squares) case}
\label{fig:matsuo}
\end{figure}

In Fig. \ref{fig:matsuo}, the removal (or addition) probabilities obtained by using 
Eqs. (6) and (15) 
are compared for the Sn isotopic chain and the two different employed pairing interactions. 
For the surface case, the nucleus $^{136}$Sn corresponds to the drip line nucleus (the two-neutron separation energy changes its sign going 
from $^{136}$Sn to $^{138}$Sn in this case).  
As can be seen in this figure, the two sets of results  can almost be superposed one to the other, as it should be if a full basis was used in both formulations.
The small difference is due to the fact that in both cases the actual calculation is made with a set of states below a certain cutoff. The cutoff
used in the quasiparticle space case cannot be easily connected to the cutoff in canonical single-particle space. Therefore, the inevitable slightly 
different cutoff choices lead to the (very) small difference.  

The surface-peaked pairing interaction systematically provides a larger probability. We notice that the transfer probability is zero in this approximation for the magic nuclei $^{100}$Sn and $^{132}$Sn due to the absence of static pairing correlation in these cases. 

\subsection{Interpretation of pair-transfer probability}

\begin{figure}[htbp]
\begin{center}
\includegraphics[width=8cm]{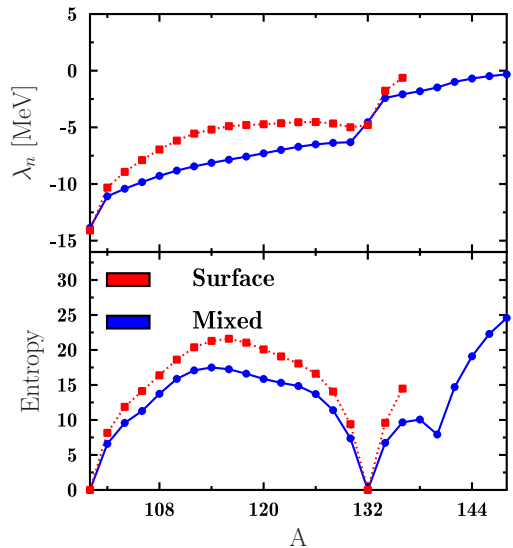}
\end{center}
\caption{Upper panel: Evolution of the neutron Fermi energy for the tin isotopic chain for mixed (filled circles) and surface
(filled squares) pairing case. Lower panel: evolution of the entropy for the same nuclei. 
}\label{fig:lambda}
\end{figure}

The canonical basis formulation of the pair transfer gives an interesting new insight in the interpretation of the transfer probability. 
Indeed, the differences between the pair-transfer probabilities which are obtained with different pairing forces are usually interpreted as a consequence 
of the radial features of the quasiparticle wave functions that 
appear in the integral of Eq. (6). One may wonder whether this is an artificial result depending on the technical details of the HFB calculations, where integrals of the components of the quasiparticle wave functions are done to evaluate the probability.  However,  
in Eq. (\ref{eq:pacan}), the radial dependence of the canonical-state wave functions is integrated out and completely disappears. 
We can thus argue that the differences in the transition probabilities are not artificial and certainly contain genuine physical  effects.  These effects that lead to a larger transfer probability for the case of a pure surface interaction with respect to the case of a mixed interaction 
could be seen 
(in the canonical basis formulation where the probability is expressed in terms of occupation numbers)
as due to a different neutron occupancy fragmentation  
around the Fermi energy. The neutron occupation numbers around the Fermi energy are more fragmented in the case of a 
 surface pairing interaction; in this latter case the Fermi energy is also closer to zero meaning that the system is less bound.
 This indicates that the last occupied states are closer to the continuum. 
This is illustrated in the upper panel of Fig. \ref{fig:lambda} where the neutron Fermi energy $\lambda_n$ is displayed as a function of $A$ for both interactions. 
 To estimate in a systematic way 
the fragmentation of the single-particle occupancies, the single-particle entropy, defined as 
\begin{eqnarray}
S & = & -k_B \sum_i \left(n_i \ln n_i + (1-n_i) \ln (1-n_i)  \right), 
\end{eqnarray}   
is shown as a function of $A$ in the lower panel of Fig. 2. The more spreading of occupation number, the larger  
should be this quantity. 
A larger fragmentation of the occupation numbers implies a more diffuse Fermi surface and thus more important surface effects. 
\begin{figure}[htbp]
\begin{center}
\includegraphics[width=8cm]{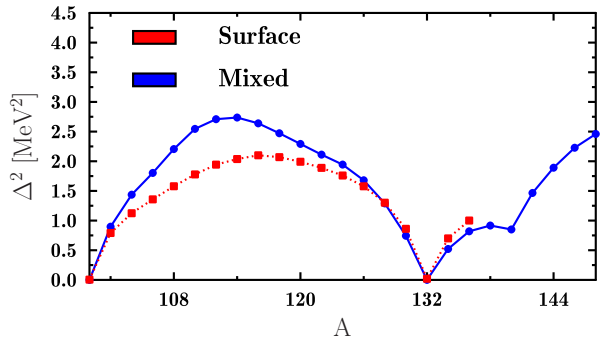}
\end{center}
\caption{Evolution of $\Delta^2$ for the tin isotopic chain for mixed (filled circles) and surface
(filled squares) pairing case. 
}\label{fig:gapentropy}
\end{figure}

This figure illustrates that the trend of the entropy is actually comparable to the trend of the strength. In both cases the results obtained with the surface-peaked pairing interaction are systematically larger than the values associated to a mixed interaction.  

It is worth spending some words about the trend of the square of the pairing gap $\Delta^2$ as a function of $A$. 
In Ref. \cite{matsuo} the authors compare $\Delta^2$ and the pair-transfer strength in Sn isotopes. They expect that these quantities should be proportional. By comparing their results, they actually observe that the two quantities have the same trend but that they are not proportional: the surface enhancement effect beyond the isotope $^{140}$Sn (in their case the drip line for the surface interaction is shifted to heavier isotopes with respect to the present calculations) is more strongly visible in the pair-transfer probability than in $\Delta^2$. 
We believe that the pair-transfer strength and the square of the pairing gap cannot be directly compared. 
These two quantities cannot scale in the same way since they are obtained by integrating 
 different functions. Furthermore, the comparison depends also on the adopted definition of the pairing gap which is not unique. 
 In Ref. \cite{matsuo} the expression 
\begin{equation}
\widetilde \Delta = \frac{\int dr \kappa(r) \Delta(r)}{\int dr \kappa(r)}
\end{equation}
is employed, where $\kappa$ is the anomalous density. 
Another definition is also currently used \cite{Dob96}, namely, 
\begin{equation}
\Delta = \frac{\int dr \rho(r) \Delta(r)}{\int dr \rho(r)},
\label{doba}
\end{equation}
where $\rho$ is the particle density.

To clarify the connection between $\Delta$ and $\widetilde \Delta$ on one side and between these quantities and the pair-transfer probabilities 
on the other side, let us go back to the canonical basis representation. The two gap expressions are 
\begin{eqnarray}
 \Delta &=& \frac{\sum_{ij} n_i \sqrt{n_j (1-n_j)} C_{ij}
}{\sum_i n_i} \nonumber \\
\widetilde \Delta &=& \frac{\sum_{ij}  \sqrt{n_i (1-n_i)} \sqrt{n_j (1-n_j)} C_{ij}
}{\sum_i  \sqrt{n_i (1-n_i)}},  
\end{eqnarray}
where 
\begin{eqnarray}
C_{ij} &=& \int dr V_{i\bar i j \bar j} |\varphi_i (r)|^2  |\varphi_j (r)|^2.
\end{eqnarray}
We see that only if $C_{ij} = 1$ for any couples of states around the Fermi energy, we have:
\begin{eqnarray}
\Delta^2 &=&  \widetilde \Delta ^2 = P^{\rm Rem/Add}_{\rm GS}(A).
\end{eqnarray}  
Such a condition is very unlikely. $\Delta$ and $\widetilde \Delta$ mix in a different way radial effects 
and fragmentation of occupation numbers showing again that these quantities 
cannot {\it a priori} directly be connected one to the other and to transfer properties.

In the case of a mixed pairing interaction the two definitions of $\Delta$ are expected to provide similar results because 
the radial localizations of $\rho(r)$ and $\kappa(r)$ are not very different.   
The case of a surface-peaked pairing interaction is however different. 
Due to the very different radial profiles of the anomalous and particle densities the two gaps may be quite different in this latter case. 
In particular, if the expression of Eq. (\ref{doba}) is used, the pairing gap can have a very  different behavior than the pair 
transition probability. This is shown in Fig. 3  where one sees that the surface-case $\Delta^2$ is lower than the volume-case 
$\Delta^2$ (inversed behavior with respect to the transfer probability) between the isotopes $^{100}$Sn and $^{132}$Sn. 

\section{Improved treatment of pair transfer}

To obtain the expression (\ref{eq:matsuo}) or (\ref{eq:pacan}), it has been necessary to assume that $| GS_A\rangle \simeq | GS_{A\pm 2} \rangle $
which is at variance with the original prescription \cite{broglia}. In this section, we discuss how to obtain, in the framework of the HFB model, a  formulation similar to that advanced in Ref. \cite{broglia}.  

\subsection{Quasiparticle formulation}

We do not adopt the approximation used in last Section and write explicitly the Bogoliubov transformations in 
 the amplitudes for the removal and addition transitions. For the removal amplitude this means,  

\begin{widetext}
\begin{eqnarray}
\nonumber
T^{\rm Rem}_{\rm  GS} (A,r) = 
\sum_{nn'} \langle{\rm GS}_{A-2} | (u^{A-2}_n({\bf r} ,-\sigma_1) \gamma_{n -\sigma_1} + (-1)^{1/2-\sigma_1} v^{*A-2}_n 
({\bf r} ,\sigma_1 ) \gamma^{\dag} _{n \sigma_1} ) \\
(u^{A}_{n'}({\bf r} , \sigma_2) \gamma_{n' \sigma_2} + (-1)^{1/2+\sigma_2} v^{*A}_{n'} 
({\bf r} , -\sigma_2 ) \gamma^{\dag} _{n' -\sigma_2}|{\rm GS}_A \rangle.  
\end{eqnarray} 
\end{widetext}

An analogous expression may be written for the addition amplitude. 
To explicitly evaluate the matrix elements of the operators $\gamma$ we have to adopt the following approximations:

\begin{widetext}
\begin{eqnarray}
\nonumber
\langle {\rm GS}_{A\pm2} | \gamma^{\dag}_{n,\sigma_1} \gamma_{n',\sigma_2} | {\rm GS}_{A} \rangle \sim  
\langle {\rm GS}_A | \gamma^{\dag}_{n,\sigma_1} \gamma_{n',\sigma_2} | {\rm GS}_A \rangle = \delta_{nn'}\delta_{\sigma_1 \sigma_2}, \\
\nonumber
 \langle {\rm GS}_{A\pm2} | \gamma_{n, \sigma_1} \gamma_{n', \sigma_2} | {\rm GS}_A \rangle \sim    
\langle {\rm GS}_A | \gamma_{n, \sigma_1} \gamma_{n', \sigma_2} | {\rm GS}_A \rangle = 0, \\
\langle {\rm GS}_{A\pm2} | \gamma^{\dag}_{n, \sigma_1} \gamma^{\dag}_{n', \sigma_2} | {\rm GS}_A \rangle \sim   
\langle {\rm GS}_A | \gamma^{\dag}_{n, \sigma_1}  \gamma^{\dag}_{n', \sigma_2}| {\rm GS}_A \rangle =0, 
\end{eqnarray}
\end{widetext}

and thus obtain for the removal and addition amplitudes

\begin{widetext}
\begin{eqnarray}
\nonumber
T^{\rm  Rem}_{\rm  GS} (A,r) 
\sim 
- \frac{1}{4\pi r^2} \sum_{nlj} (2j+1) u^{A-2}_{nlj} (r) v^{A}_{nlj} (r), \\
T^{\rm  Add}_{\rm  GS} (A,r) 
\sim 
- \frac{1}{4\pi r^2} \sum_{nlj} (2j+1) u^{A}_{nlj} (r) v^{A+2}_{nlj} (r).
\end{eqnarray} 
\end{widetext}

The improvement with respect to the previous derivation is achieved by actually making the approximation $|GS_A \rangle \sim |GS_{A\pm 2} \rangle $ on the matrix elements of the creation/annihilation operators $\gamma^{\dag},\gamma$. 
That is, the approximation is done after having written and explicitly applied 
the Bogoliubov transformations in the expression of the transition amplitudes.This leads to expressions where the wave functions of the two nuclei appear as originally used in ref. \cite{broglia}. 

The pair-transfer strengths are given by 

\begin{equation}
P^{\rm  Rem}_{\rm  GS}(A) = \left| \int dr \sum_{nlj} (2j+1) u^{A-2}_{nlj} (r) v^{A}_{nlj} (r) \right|^2,   \label{eq:marcella}
\end{equation} 
\begin{equation}
P^{\rm  Add}_{\rm  GS}(A) = \left| \int dr \sum_{nlj} (2j+1) u^{A}_{nlj} (r) v^{A+2}_{nlj} (r) \right|^2.   
\end{equation} 
One notices also that 
\begin{equation}
P^{\rm  Rem}_{\rm  GS} (A+2) =   P^{\rm  Add}_{\rm GS} (A).  \label{eq:addrem}
\end{equation}

\subsection{Canonical formalism}

Similarly to what is done above in the quasiparticle formulation, a better 
approximation can be obtained also in the canonical formalism. Assuming that the canonical states do not change 
too much between the nucleus $A$ and $A \pm 2$ (no core polarization due to the addition
or removal of two nucleons) the improved formula are

\begin{widetext}
\begin{eqnarray}
T^{\rm Add}_{\rm GS} (A, {\bf r})& \simeq & - \sum_{i>0} \langle {\rm GS}_{A+2} | a^\dagger_i  a^\dagger_{\bar i} 
| {\rm GS}_A \rangle 
\left(|\varphi_i ({\bf r}, \uparrow) |^2  + |\varphi_i ({\bf r}, \downarrow) |^2 \right) \nonumber \\
&=& \sum_{i>0} \sqrt{n^{A+2}_i (1 - n^{A}_i)}  \left(|\varphi_i ({\bf r}, \uparrow) |^2  + |\varphi_i ({\bf r}, 
\downarrow) |^2 \right) 
\nonumber 
\end{eqnarray} 
and 
\begin{equation}
T^{\rm Rem}_{\rm GS} (A, {\bf r}) = - \sum_{i>0} \sqrt{ (1 - n^{A-2}_i) n^{A}_i}  \left(|\varphi_i ({\bf r}, \uparrow) |^2  + |\varphi_i ({\bf r}, \downarrow) |^2 \right). 
\label{eq:cortrans} 
\end{equation} 
\end{widetext}       
Accordingly, the two-particle addition and removal probabilities 
now read 
\begin{widetext}
\begin{eqnarray}
P^{\rm Add}_{\rm GS} (A) &=&   \left|  \sum_i \sqrt{n^{A+2}_i (1 - n^{A}_i)} \right|^2 
=  
\left|  \sum_i (2j+1) \sqrt{n^{A+2}_{nlj} (1 - n^{A}_{nlj} )} \right|^2 , \label{eq:padd2} \\
P^{\rm Rem}_{\rm GS} (A) &=&  \left|  \sum_i \sqrt{(1 - n^{A-2}_i) n^{A}_i} \right|^2 
=\left|  \sum_i (2j+1)  \sqrt{(1 - n^{A-2}_{nlj} ) n^{A}_{nlj} } \right|^2.  \label{eq:rem2}
\end{eqnarray}  
\end{widetext}
It should be noted that the approximation made to obtain the above expressions slightly differs from 
the one  used in the previous subsection and differences between the two sets of results may actually be expected within the improved formalism. 

\subsection{Results and discussion}

\begin{figure}[htb]
\begin{center}
\includegraphics[width=8cm]{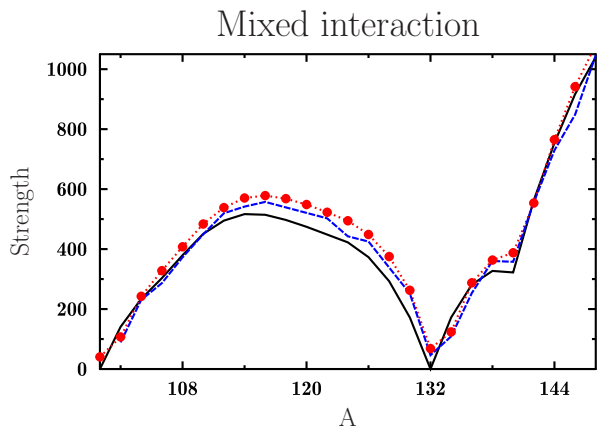}
\end{center}
\caption{
Comparison of the removal probability obtained with the mixed 
pairing case using Eq. (\ref{eq:matsuo}) (solid line) and the improved expressions
given by Eqs. (\ref{eq:marcella}) (dashed curve) and 
(\ref{eq:rem2}) (filled circles). 
}
\label{fig:impmixed}
\end{figure}

\begin{figure}[htb]
\begin{center}
\includegraphics[width=8cm]{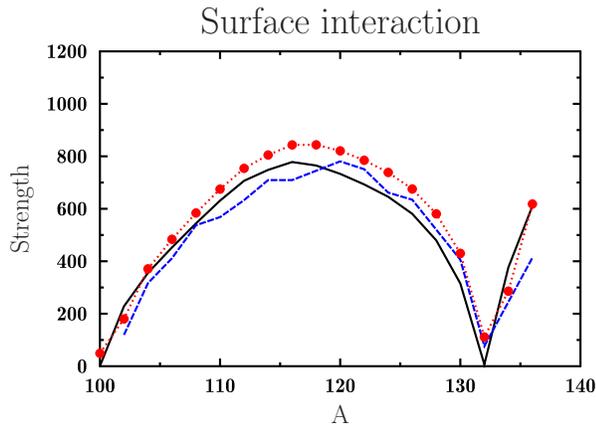}
\end{center}
\caption{Same as in Fig. \ref{fig:impmixed} but for the case of a surface-peaked pairing 
interaction.
}
\label{fig:impsurf}
\end{figure}

Since the difference between removal and addition probabilities is just a shift of nuclei (see Eq. (\ref{eq:addrem})), we consider in what follows only removal probabilities. 
In Figs. \ref{fig:impmixed} and \ref{fig:impsurf} the results obtained with the improved expressions for the removal strength 
are presented for the mixed and the surface interaction, respectively.  
The canonical basis results (filled circles) are compared with the quasiparticle results (dashed line) and with the results obtained 
with Eq. (\ref{eq:matsuo}) (solid line). 
In figure \ref{fig:impmixed}, we see that 
 the canonical and quasiparticle improved forms give
 similar results.  In particular, these new forms of the pair-transfer probability lead to non-zero 
values for magic nuclei. It is also worth mentioning that the probability is enhanced in the mid-shell with respect to the corresponding value  obtained using Eq. (\ref{eq:matsuo}).  

The case of Figure \ref{fig:impsurf}  is different. 
We observe some irregularities in the trend of the results which have been obtained with Eq. (\ref{eq:marcella}). These irregularities are related to some specific features of the HFB calculations in coordinate representation. It may happen that 
the fragmentation of the occupations among the discretized quasiparticle states (box boundary conditions with a box radius equal to 20 fm) is not the same in the nucleus $A\pm2$ and in the nucleus $A$. When this is the case, some irregularities appear in the transfer strength calculated by making products of functions belonging to the different nuclei $A$ and $A\pm2$. This situation does not occur when one uses the canonical basis, that displays a smoother behavior when moving from one system to the neighbor one. There irregularities are much more pronounced
in the case of surface-peaked interaction. These results indicate that the improved treatment of pair transfer discussed here should be 
done using a discrete basis instead of the coordinate representation to avoid the irregularities.   

We also observe that the differences between the results obtained with the canonical and the quasiparticle formulations are more pronounced in the case of a surface-peaked interaction (Fig. 5). This may be related to the fact that the artificial irregularities of the quasiparticle results are more important in the case of a surface pairing interaction (the fragmentation of the occupation numbers may vary more strongly from the nucleus $A$ to the nucleus $A\pm2$ in this case). The slightly different approximation adopted for the ground states of the two nuclei in the two derivations  also accounts for these differences. 

\section{Effect of particle number conservation on pair transfer probability}

The quasiparticle states which are generally used to estimate pair-transfer probabilities are not eigenstates of the 
particle number operator. In particular, a state $| {\rm GS}_A \rangle $  contains not only components with  
the correct number of particles, but also components with particle numbers $A\pm 2$, $A \pm 4$, ... 
inducing spurious contributions to the probabilities. These contributions can be exactly removed using 
projection techniques.  In the following, it is assumed that the ground state of a system with particle number $A$ 
is written as 
\begin{eqnarray}
| {\rm GS}_A \rangle   \simeq | A \rangle =    P^A \prod_{i>0} \left( u^A_i + v^A_i a^\dagger_i a^\dagger_{\bar i} \right) | 0 \rangle, \label{eq:pbcs} 
\end{eqnarray}
where $P^A$ denotes the projector onto the number of particles $A$. The projected state has the particularity that it has the same canonical basis as the original quasiparticle vacuum from which it is constructed. As shown in Appendix  \ref{sec:appA}, 
the transition probability accounting for particle number projection can now be approximately estimated by replacing 
occupation numbers of the quasi-particle state $ n^A_i$ in Eqs. (\ref{eq:padd2}) and (\ref{eq:rem2}) 
by the new equivalent occupation numbers, denoted by $\bar n^A_i$, in the projected state, i.e.:
\begin{eqnarray}
{\tilde P}^{\rm Add}_{\rm GS} &=&   \left|  \sum_i \sqrt{\bar n^{A+2}_i (1 - \bar n^{A}_i)} \right|^2, 
\label{eq:padd2proj} \\
{\tilde P}^{\rm Rem}_{\rm GS} &=&  \left|  \sum_i \sqrt{(1 - \bar n^{A-2}_i) \bar n^{A}_i} \right|^2 . \label{eq:rem2proj}
\end{eqnarray}  

\begin{figure}[htbp]
\begin{center}
\includegraphics[width=8cm]{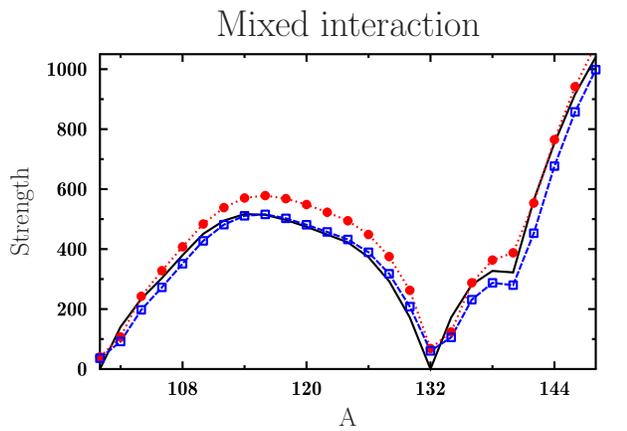}
\end{center}
\caption{Removal transfer probability obtained accounting for particle number conservation (open square) and compared 
to the improved (Sec. IV) (filled circles) and more approximated (Sec. III) expressions (solid line) for the mixed interaction.}
\label{fig:projmix}
\end{figure}

\begin{figure}[htbp]
\begin{center}
\includegraphics[width=8cm]{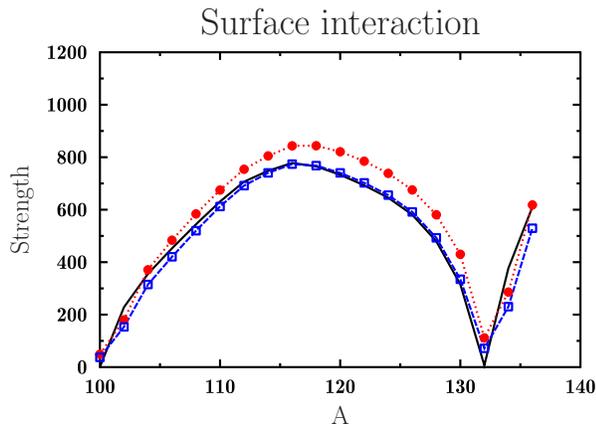}
\end{center}
\caption{Same as figure \ref{fig:projmix} for the case of pure surface pairing 
interaction.}
\label{fig:projsurf}
\end{figure}

In Figs. \ref{fig:projmix} and \ref{fig:projsurf}, the effect of particle number conservation on the estimation of the transition probabilities is illustrated 
for mixed and pure surface pairing, respectively. To calculate the occupation number of the projected state, standard
gauge angle integration technique has been used with a Fomenko \cite{Fom70} discretization using 199 points 
(see for instance Ref. \cite{Ben09}). 

In these figures, we clearly see a very interesting and unexpected effect: at mid-shells, the projection actually tends to cancel out the effect of using the occupation numbers  
of the $A$ and $A\pm 2$ nuclei in the improved formulation of last Section.  The pairing strength gets again closer to the probability obtained with the approximation of Ref. \cite{matsuo} and of Sec. III.  
However, close to magicity, projection has little effect and the results remain unchanged with respect to the improved treatment of the transfer probability. We mention that in Ref. \cite{matsuo}, in order to have a non zero probability at shell closures, the transfer probability is calculated in these cases with the particle-particle random-phase approximation. 
Non-zero values 
are obtained here within  a different approach based on a unified model suited to treat all the nuclei, both at shell closures and in mid-shell regions.  However, the strength at shell closure is much lower here compared to Ref. \cite{matsuo}.




 
\section{Conclusions}

Starting from an approximate derivation of the pair-transfer probability within the HFB approach, we have investigated its range of validity. 
Two different pairing interactions with a different surface/volume mixing are used. 
 To get a deeper and complementary physical insight 
we have formulated the same problem also in the canonical basis representation. 
In this case, the radial dependence of the wave functions is completely integrated out 
and disappears in the transfer probability.  
The same surface effects which are due to the integrated radial profiles of the contributing wave functions in the quasiparticle representation still exist in the canonical basis case as a consequence of the different fragmentation of the single-particle canonical occupancies. 
The transfer probability is actually connected to the diffuseness/ fragmentation
of the single-particle state occupancies close to the Fermi energy.

The possibility to improve the description of pair transfer probabilities in a mean-field model is then discussed. Two 
improvements are introduced: (i) the use of two different quasi-particle vacuum  for the initial and final ground states; (ii)
the possibility to exactly restore the proper number of particles in the entrance channel. It is shown that these refinements 
are important especially close to the magicity where a non-zero probability is found contrary to the simplest method. 
It turns out that the enhancement observed in the mid-shell by taking two different ground states is partially compensated 
by the particle number restoration. 
We can thus finally conclude that the simple formula used in Ref. \cite{matsuo}, although approximated, can still  provide good results at mid-shells. 
We remind however that the projection is performed here {\it a posteriori}.  
If variation after projection is made, additional correlations 
  are expected to appear especially close to magic numbers \cite{hupin}. 
This open new perspectives and could be an interesting subject for future investigations. 

\newpage 

\appendix

\section{Transfer probability for states with good particles number}
\label{sec:appA}

In this section, we consider two states denoted by $|A \rangle$ and $| A+2 \rangle$ obtained 
by projecting quasi-particle states onto good particle numbers, i.e. \footnote{Note that, consistently with the 
approximation made in section \ref{sec:can}, it is implicitly assumed that this many-body states shares the same canonical 
single-particle basis.}$^,$\footnote{Here, the convention $| A \rangle_B$ is taken for the state:
\begin{eqnarray} 
| A \rangle_B & \equiv & P^A \prod_{i} \left( 1 + x^B_i a^\dagger_i a^\dagger_{\bar i} \right) | 0 \rangle 
\end{eqnarray} 
where the $B$ refers to the fact that the quasi-particle vacuum has been obtained with the constraint that 
the number of particle is fixed to B in average.  
}
\begin{eqnarray} 
| A \rangle_A & \equiv & P^A \prod_{i} \left( 1 + x^A_i a^\dagger_i a^\dagger_{\bar i} \right) | 0 \rangle \label{eq:pa} \\
| A + 2 \rangle_{A+2} & \equiv & P^{A+2} \prod_{i} \left( 1 + x^{A+2}_i a^\dagger_i a^\dagger_{\bar i} \right) | 0 \rangle
\label{eq:pa2}
\end{eqnarray} 
where, we have introduced the coefficients $x^A_i = v^A_i /u^A_i$ and $x^{A+2}_i = v^{A+2}_i /u^{A+2}_i$. 
The addition transfer probability accounting for the particle number conservation requires to estimate 
the  quantity ${}_{A+2}  \langle A + 2 | a^\dagger_i a^\dagger_{\bar i}  | A \rangle_A$. An approximate form 
is derived below. 
\subsection{Some properties of projected states}

The projection has the effect to select the component with good particle number in the quasiparticle vacuum leading to:   
\begin{widetext}

\begin{eqnarray} 
| A \rangle_A & \equiv & 
\frac{1}{\sqrt{N!}}
 \left(\sum_i x^A_i a^\dagger_i a^\dagger_{\bar i} \right)^N  | 0 \rangle \\
| A + 2 \rangle_{A+2} & \equiv & 
\frac{1}{\sqrt{(N+1)!}}  
\left(\sum_i x^{A+2}_i a^\dagger_i a^\dagger_{\bar i} \right)^{N+1}  | 
0 \rangle
\end{eqnarray} 
where $N$ and $N+1$ denotes the number of pairs.
The properties of projected state have been recently reviewed in Refs. \cite{Lac10,Hup11}
and some of them will be recalled below. For instance, it has been shown that several recurrence relation exists to manipulate 
this state. Indeed, by developing the power in previous expressions 
and noting that $(a^\dagger_i a^\dagger_{\bar i})^2 = 0$, we deduce:
\begin{eqnarray}
| A \rangle_A &=&  
\frac{1}{\sqrt{N!}} 
\left\{ \left(\sum_{j\neq i} x^A_j a^\dagger_j a^\dagger_{\bar j}\right)^N + N x^A_i a^\dagger_i a^
\dagger_{\bar i} 
\left(\sum_{j\neq i} x^A_j a^\dagger_j a^\dagger_{\bar j} \right)^{N-1}  \right \}  | 0 \rangle \nonumber \\
&\equiv& | A : i \rangle_A + \sqrt{N} x^A_i a^\dagger_i a^\dagger_{\bar i}    | A-2 : i \rangle_A
\end{eqnarray}
from which we obtain the recurrence relation on the overlaps:
\begin{eqnarray}
 \langle A : i  | A : i \rangle_A  =  \langle A   | A  \rangle_A   -  N |x^A_i|^2 \langle A - 2 : i  | A - 2 : i \rangle_A . \label{eq:recover}
\end{eqnarray} 
By definition, all the states $|  . : i \rangle_B$ do not contain the pairs $a^\dagger_i a^\dagger_{\bar i} $. Accordingly, we have:
\begin{eqnarray}
a^\dagger_i a^\dagger_{\bar i}  | A \rangle_A & = &  | A : i \rangle_A , \\
_{A+2} \langle A + 2 | a^\dagger_i a^\dagger_{\bar i} & = &   \sqrt{(N+1)}  x^{A+2*}_i  {}_{A+2} \langle A : i  |.
\end{eqnarray} 
With these expression, we finally see that we have the relationship:
\begin{eqnarray}
{}_{A+2}  \langle A + 2 | a^\dagger_i a^\dagger_{\bar i}  | A \rangle_A  & = &  \sqrt{(N+1)}  x^{A+2*}_i  
{}_{A+2} \langle A : i  | A : i \rangle_{A} .
\end{eqnarray}

\subsection{Approximate form of the transition probability}

States defined by Eqs. (\ref{eq:pa} - \ref{eq:pa2}) are non-normalized states. To estimate transition density, one should 
use the following normalized states instead:
  \begin{eqnarray}
 | \widetilde A  \rangle_A  &=& \frac{1}{\sqrt{ \langle A   | A  \rangle_A}}  | A  \rangle_A,  \\
 | \widetilde {A + 2}  \rangle{}_{A+2} & =& \frac{1}{\sqrt{ \langle A  + 2  | A + 2 \rangle_{A+2}}}  | A + 2  \rangle_{A+2},
\end{eqnarray} 
leading to:
\begin{eqnarray}
{}_{A+2}\langle \widetilde {A + 2} | a^\dagger_i a^\dagger_{\bar i}  | \widetilde A \rangle{}_{A} & = & \frac{\sqrt{(N+1)}  x^{A+2*}_i }
{\sqrt{\langle A   | A  \rangle_A \langle A +2  | A +2 \rangle_{A+2}}}  {}_{A+2} \langle A : i  | A : i \rangle {}_{A}.
\label{eq:trans1} \nonumber
\end{eqnarray}
Assuming that
\begin{eqnarray}
{}_{A+2} \langle A : i  | A : i \rangle_{A}   & \simeq &  {}_{A+2} \langle A : i  | A : i \rangle_{A+2} \simeq   
{}_{A} \langle A : i  | A : i \rangle_{A}
\end{eqnarray}
and using the Eq. (\ref{eq:recover}), two approximate forms of the transition can be obtained:
\begin{eqnarray}
{}_{A+2}\langle \widetilde {A + 2} | a^\dagger_i a^\dagger_{\bar i}  | \widetilde A \rangle{}_{A} & \simeq & \frac{\sqrt{(N+1)}  x^{A+2*}_i }
{\sqrt{\langle A   | A  \rangle_A \langle A +2  | A +2 \rangle_{A+2}}}  {}_{A+2} \langle A : i  | A : i \rangle {}_{A+2}
\label{eq:trans1prime} \nonumber
\end{eqnarray}
and 
\begin{eqnarray}
{}_{A+2}\langle \widetilde {A + 2} | a^\dagger_i a^\dagger_{\bar i}  | \widetilde A \rangle_A & \simeq & \frac{\sqrt{(N+1)}  x^{A+2*}_i }
{\sqrt{\langle A   | A  \rangle_A \langle A +2  | A +2 \rangle}_{A+2}} \left( \langle A   | A  \rangle_A   -  N |x^A_i|^2 \langle A - 2 : i  | A - 2 : i \rangle_A  \right).
\label{eq:trans2} \nonumber
\end{eqnarray} 
Combining these two expressions, it could be checked that:
\begin{eqnarray}
| {}_{A+2} \langle \widetilde {A + 2} | a^\dagger_i a^\dagger_{\bar i}  | \widetilde A \rangle _A|^2 & \simeq & (N+1) | x^{A+2*}_i |^2 
\frac{\langle A  : i  | A : i \rangle_{A+2}}{\langle A +2  | A +2 \rangle_{A+2} } \times \left(1 - N |x^A_i|^2 \frac{\langle A - 2 : i  | A - 2 : i \rangle_A}{\langle A   | A  \rangle_A} \right) \nonumber
\end{eqnarray}
We recognize in this expression nothing but the occupation probabilities of the state $i$ 
in the projected state respectively  with A and A+2 particles \cite{Lac10} given by:
\begin{eqnarray}
\bar n^A_i & = &  N |x^A_i|^2 \frac{\langle A - 2 : i  | A - 2 : i \rangle_A  }{ \langle A   | A  \rangle_A }
\end{eqnarray}  
and 
\begin{eqnarray}
\bar n^{A+2}_i & = &  (N+1) |x^{A+2}_i|^2 \frac{\langle A  : i  | A : i \rangle_{A+2}  }{ \langle A+2   | A+2  \rangle_{A+2} }.
\end{eqnarray}
Altogether, we obtain that the  transition amplitude between two states with good particle number can be approximated 
by 
\begin{eqnarray}
{}_{A+2} \langle \widetilde {A + 2} | a^\dagger_i a^\dagger_{\bar i}  | \widetilde A \rangle _A  & \simeq & 
\sqrt{\bar n^{A+2}_i  (1- \bar n^A_i) }
\end{eqnarray}  
that is nothing but the same expression as the one obtained in the case of non-projected state except 
that the occupation numbers entering here are those associated with the projected states.
\end{widetext}

\end{document}